%% file: XL_RIS_SecureComm.tex
\documentclass[lettersize,journal]{IEEEtran}
\usepackage{amsmath,amssymb,amsfonts}
\usepackage[utf8]{inputenc}
\usepackage{algorithmic}
\usepackage{algorithm}
\usepackage{array}
\usepackage[caption=false,font=scriptsize,labelfont=sf,textfont=sf]{subfig}
\usepackage{textcomp}
\usepackage{stfloats}
\usepackage{url}
\usepackage{verbatim}
\usepackage{graphicx}
\usepackage{cite}
\usepackage{amssymb}
\usepackage{setspace}
\usepackage{lineno} 
\usepackage{cases}
\usepackage{subfloat}
\usepackage{float}  
\usepackage{titlesec}
\usepackage{textcomp}
\usepackage{bm,bbold}
\usepackage{xcolor}
\usepackage{siunitx}
\input{sym_marco.tex}

\begin{document}

\title{Beamforming Optimization for Extremely Large-Scale RIS-Aided Near-Field Secure Communications}

\author{Xiaotong Xu,
		Qian Zhang,
		Yunxiao Li,
		Xuejun Cheng,
		Meihui Liu,
		and Ju Liu
\thanks{Qian Zhang is with School of Information Science and Engineering, Shandong University, Qingdao 266237, China, and with School of Electrical and Electronic Engineering, Nanyang Technological University, Singapore.}
\thanks{Xiaotong Xu, Yunxiao Li, Xuejun Cheng, Meihui Liu, and Ju Liu are with School of Information Science and Engineering, Shandong University, Qingdao 266237, China.}

}

\maketitle

\begin{abstract}
This paper studies an extremely large-scale reconfigurable intelligent surface (XL-RIS)-aided near-field physical layer security (PLS) communication system, aiming to maximize the secrecy rate by jointly optimizing precoding vector at the BS and the reflection coefficient matrix at the XL-RIS. Artifi-cial jamming was introduced to further enhance communication security. To solve the non-convex secrecy rate problem, an alternate optimization-based algorithm is adopted to decompose it into two sub-problems. Specifically, when optimizing the transmit beamformer at the BS, the non-convex prob-lem is transformed into a convex one through the weighted minimum mean-square error and the successive convex approximation-based algorithms. For the optimization problem of the XL-RIS phase-shifting matrix, a low-complexity alternating direction method of multipliers-based algorithm is employed to enhance the flexibility of the design. The proposed algorithm is capable of accommodating discrete phase optimization for the XL-RIS, thus better aligning with practical system requirements. Simulation results demonstrate that when the eavesdropper reside in the same direction as the legitimate user and is located closer to the XL-RIS, the proposed scheme in this paper can still ensure the secure communication.

\end{abstract}


\section{Introduction} 
Due to the broadcast characteristic, wireless communication is vulnerable to infor-mation leakage. As a promising technology for 6G, the reconfigurable intelligent surface (RIS) has the potential to improve the overall communication quality~\cite{Gong2020IRS,Zhang2025SIM_ISAC,Zhang2025BDRIS}.

When applied to physical layer security (PLS), RIS enhances the desired signal power at legitimate users while suppressing the power at eavesdroppers by passive beamforming. This enables secure and energy-efficient transmission~\cite{Xu2025IRS,Zhang2023RIS_NOMA}. However, RIS’s performance is limited by the double fading effect. Considering that the array gain of RIS scales quadratically with the number of elements~\cite{Sun2022RIS}, employing a large number of low-cost reflecting elements can mitigate this degradation. The concept of extremely large-scale RIS (XL-RIS) has been proposed~\cite{Cui2023NFMIMO,Zhang2026XL_RIS}. As the aperture of the RIS increases, communication tends to occur in the near-field region.

Unlike the far-field planar wave model, near-field spherical-wave model adopts an accurate electromagnetic wave propagation which incorporates both angular and radial information~\cite{Zhang2024PLS,Zhang2023RIS_NOMA}. The authors in~\cite{Zhang2024PLS} demonstrated that beamfocusing in near-field scenarios can achieve secure communication under angular alignment of legit-imate and illegitimate users. Authors in~\cite{Liu2024RIS} studied XL-RIS aided near-field covert communication could achieve a higher level of covertness compared to far-field setups.
To strengthen the secure transmission, several PLS techniques have been pro-posed, such as artificial jamming~\cite{Wang2022IRS_NOMA}, beamforming~\cite{Zhang2023RIS_NOMA}, and cooperative relaying~\cite{Hu2019FD}. 

However, research on PLS in the near-field communication enabled by XL-RIS remains limited. This paper aims to fill this gap by investigating joint beamforming and phase shift matrix optimization in the XL-RIS-aided near-field secure commu-nication. Specifically, we formulate a secrecy rate maximization problem under constraints on transmit power at the BS, quality of service (QoS), successive inter-ference cancellation (SIC), and constant modulus for XL-RIS reflection elements. To tackle the non-convex problem, we propose an alternating optimization (AO)-based algorithm. For the beamformer design, a weighted minimum mean square error (WMMSE)-based algorithm is proposed. For the XL-RIS’s reflection coefficient matrix design, a low-complexity alternating direction method of multipliers (ADMM)-based algorithm is used. To account for practical hardware, we incorpo-rate a discrete phase shift optimization scheme based on line search, making the proposed design more feasible for practical deployment. Simulation results show that the proposed scheme can improve secrecy performance compared to baseline schemes.

{\it{Notation:}} 
$(\cdot)^{*}$, $(\cdot)^{\rm T}$ and $(\cdot)^{\rm H}$ denote the conjugate, transpose, and conjugate transpose operation, matrix inverse operation, respectively.
$\mathbb{C}$ and $\mathbb{R}$ represent complex and real space, respectively.
${\cal CN}(\cdot,\cdot)$ denotes the circularly symmetric complex Gaussian distribution.
${\cal R}\{\cdot\}$ and ${\cal I}\{\cdot\}$ denote the real and imaginary parts, respectively.
${\rm vec}(\cdot)$ and ${\rm Tr}\left\{\cdot \right\}$ represent the vectorization and trace of the matrix, respectively.
$j$ represents the imaginary unit.
$\bm 1_N$ is a full-one vector with $N$ elements.
$\bold{I}_K$ denotes a $K \times K$ identity matrix.
$[\bz]_n$ denotes the $n$-th entry of the vector $\bz$.
$\left\| \cdot \right\|_{*}$, $\|\cdot\|_2$, and $|\cdot|$ denote the unclear norm, Euclidean norm, and absolute value, respectively.
$\Pi_{{\cal C}}$ denotes a projection operation.

\begin{figure}[t]	
	\centering \includegraphics[width=\linewidth]{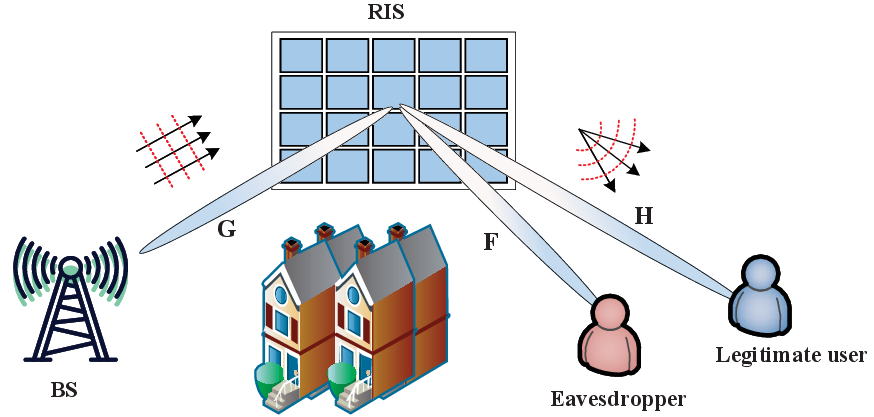}
	\caption{XL-RIS-aided downlink secure communication system model.}
	\label{fig:System_Model}
\end{figure}

\section{System Model}
In this section, we consider a downlink multiple-input single-output (MISO) secure communication system assisted by an XL-RIS, as illustrated in Fig.~\ref{fig:System_Model}. 
The base station (BS) is equipped with $M$ antennas. 
The XL-RIS is a uniform planar array (UPA) with $N=N_1 \times N_2$ reflecting elements. 
Both the legitimate user and the eaves-dropper are assumed to have no direct links with the BS due to blockage. 
The phase shift matrix of the XL-RIS is given by $\bTheta = \diag\left( e^{j\theta_1},\dots,e^{j\theta_N} \right)$, where $\theta_n\in [-\pi,\pi)]$. 
The boundary between far-field and near-field regions is defined by the Rayleigh distance $d=2D^2/\lambda$, where $D$ is the antenna aperture and $\lambda$ is the carrier wavelength.
Following practical deployment considerations~\cite{Cheng2024RISNF}, the XL-RIS is assumed to be closer to the user side to maximize performance. Thus, the BS is placed in the far-field region of the XL-RIS, while the receivers lie in the near-field region. 

We assume that the legitimate user’s CSI can be obtained by downlink pilot signals from the BS~\cite{Wang2021IRS}. Moreover, assuming that the BS can acquire the eavesdrop-per’s CSI: i) when the eavesdropper pretends to be legitimate, the CSI is available to the BS due to legal interaction. ii) when the eavesdropper is malicious, radiometers are commonly used to monitor the communication between the BS and user. In practice, nearly all radiometers are superheterodyne receivers whose signal leakage is unavoidable.

The far-field channel $\bG\in \mathbb{C}^{N\times M}$ between the BS and XL-RIS can be modeled as
\begin{equation}
	\begin{split}
		\bG = \sum_{i=1}^{L_1} \alpha_i \bb(\varphi_i,\eta_i)\ba^{\rm H}(\gamma_i),
	\end{split}
\end{equation}
where $L_1$ denotes the number of propagation paths, $\alpha_i$ is the complex gain of the $i$-th path.
The array response vectors at the BS and XL-RIS are given by
\begin{equation}
	\begin{split}
		\ba(\gamma) = \frac{1}{\sqrt{M}} \left[e^{j \frac{2\pi d}{\lambda}m \sin\psi } \right]^{\rm T}_{m\in L(M)},
	\end{split}
\end{equation}
\begin{equation}
	\begin{split}
		\bb(\varphi,\eta) = &\frac{1}{\sqrt{N}} \left[e^{j \frac{2\pi d}{\lambda}n_1 \sin\varphi \cos\eta } \right]^{\rm T}_{n_1\in L(N_1)} \\
		&\otimes \left[e^{j \frac{2\pi d}{\lambda}n_2 \sin\eta }  \right]^{\rm T}_{n_2\in L(N_2)},
	\end{split}
\end{equation}
where $L(n)=\left\{0,1,\dots,n-1\right\}$ is the index set.
$d$ is the antenna spacing.  
$\psi$ is the BS azimuth angle.
$\varphi$ and $\eta$ represent azimuth and elevation angles of the XL-RIS, respectively.

The near-field channels between XL-RIS and the receivers adopt a spherical wave model. 
We take the channel $\bh$ between XL-RIS and the legitimate user as an example.
\begin{equation}
	\begin{split}
		\bh = \beta_1 \bb(\br_{re}),
	\end{split}
\end{equation}
where $\bh\in \mathbb{C}^{N\times 1} $, $\beta_1$ is the channel gain. 
The steering vector $\bb(\br_{re})$ can be given in~\eqref{steer_vector}, where $D_{\rm re}(n_1,n_2)$ is the distance between XL-RIS $(n_1,n_2)$-th element and the user.
\begin{figure*}
	\begin{equation} \label{steer_vector}
		\begin{split}
			\bb(\br_{re}) = \frac{1}{\sqrt{N}} \left[e^{-j\frac{2\pi}{\lambda}D_{re}(1,1)}, \dots, e^{-j\frac{2\pi}{\lambda}D_{re}(1,N_2)}  ,\dots,e^{-j\frac{2\pi}{\lambda}D_{re}(N_1,1)},\dots,e^{-j\frac{2\pi}{\lambda}D_{re}(N_1,N_2)}\right]^{\rm T}.
		\end{split}
	\end{equation}
\end{figure*}

To guarantee the secure transmission, we assume that artificial jamming is generated with the signal at the BS simultaneously. 
The transmit signal can be expressed as $\bx=\bw s + \bw_{\rm jam} z$,where $\bw\in \mathbb{C}^{M\times 1}$ and $\bw_{\rm jam}\in \mathbb{C}^{M\times 1}$ denote the precoding vector and artificial jamming vector, respectively.
$s\sim {\cal CN}(0,1)$ is the data symbols sent by the BS. 
$z\sim {\cal CN}(0,1)$  is the jamming signal. 
The received signals can be given by
\begin{equation}
	\begin{split}
		y = \bh_u^{\rm  H}\bx + n,~~~y_e = \bh_e^{\rm H} \bx+n_e,
	\end{split}
\end{equation}
where $\bh_u^{\rm H} = \bh^{\rm H}\bTheta \bG$; $\bh_e^{\rm H} = \bm f^{\rm H}\bTheta \bG$; $n\sim {\cal CN}(0,\sigma^2)$ and $n_e \sim {\cal CN}(0,\sigma_e^2)$ denote the additive white Gaussian noise (AWGN). 
To mitigate the adverse impact of artificial jamming on the legitimate user with SIC, jamming power must satisfy $\left|\bh_u^{\rm H}\bw \right|^2 \leq \left|\bh_u^{\rm H}\bw_{\rm jam} \right|^2 $. 
Therefore, the signal-to-interference-plus-noise ratio (SINR) at the legitimate user is ${\rm SINR}_u = \left|\bh_u^{\rm H}\bw \right|^2/\sigma^2 $. 
The rate of the legitimate user and the eavesdropper can be expressed as $R_u = \log(1+{\rm SINR}_u)$ and $R_e = \log\left(1+\left|\bh_e^{\rm H}\bw \right|^2/\left(\left|\bh_e^{\rm H}\bw_{\rm jam} \right|^2 + \sigma_e^2\right)\right)$.
The secrecy rate can be expressed as
\begin{equation}
	\begin{split}
		R = \left[R_u - R_e \right]^{+} = \max\left(0,R_u-R_e \right).
	\end{split}
\end{equation}

\section{Algorithm Design for Secure Communication Systems}
\subsection{Problem Formulation}
We aim to maximize the secrecy rate by jointly optimizing the BS precoding vector and the XL-RIS phase shift matrix. The optimization problem can be formulated as
\begin{equation}
	\begin{split}
		\textbf{P1}: \max_{\bw,\bw_{\rm jam},\btheta}&~g(\bw,\btheta) = \left[R_u - R_e \right]^{+} \\
		{\rm s.t.}&~{\cal C}_{\rm BS}: \left\| \bw \right\|_2^2 + \left\| \bw_{\rm jam} \right\|_2^2 \leq P_{\rm max}, \\
		&~{\cal C}_{\rm th}: {\rm SINR}_u \geq r_{\rm th}, \\
		&~{\cal C}_{\rm SIC}: \left| \bh_u^{\rm H}  \bw_{\rm jam} \right|^2 \geq \left| \bh_u^{\rm H} \bw \right|^2, \\
		&~{\cal C}_{\rm RIS}: |\theta_n|=1, ~n=1,2,\dots,N,
	\end{split}
\end{equation}
where ${\cal C}_{\rm BS}$ is the transmit power constraint at the BS; ${\cal C}_{\rm th}$ is the QoS of the legitimate user; ${\cal C}_{\rm SIC}$ is the decoding constraint; ${\cal C}_{\rm RIS}$ enforces the constant modulus of XL-RIS. 
Due to the non-convex problem and coupling optimization variables, \textbf{P1} is difficult to solve. Thus, we propose an AO algorithm that decomposes\textbf{P1} into two subproblems.

\subsection{Optimization of BS Precoding Vectors}
Given the phase shift matrix $\bTheta$ of the XL-RIS, \textbf{P1} can be reformulated as
\begin{equation}
	\begin{split}
	\textbf{P2}: \max_{\bw,\bw_{\rm jam}}~g(\bw)= \left[R_u - R_e \right]^{+} ~~~{\rm s.t.}~{\cal C}_{\rm BS},{\cal C}_{\rm th},{\cal C}_{\rm SIC}.
	\end{split}
\end{equation}

To tackle the non-convex problem, the WMMSE method is adopted 
\begin{equation}
	\begin{split}
		\min_{\bw,\bw_{\rm jam}}~g_u(\bw,u_u,\rho_u) + g_e(\bw,\bw_{\rm jam},u_e,\rho_e)~{\rm s.t.}{\cal C}_{\rm BS},{\cal C}_{\rm th},{\cal C}_{\rm SIC},
	\end{split}
\end{equation}
where 
$$g_u(\bw,u_u,\rho_u) = \rho_u F_u(\bw,u_u) - \log(\rho_u),$$
$$g_e(\bw,\bw_{\rm jam},u_e,\rho_e) = \rho_e F_e(\bw,\bw_{\rm jam},u_e) - \log(\rho_e),$$
$$F_u(\bw,u_u)=|u_u|^2 \left(\left|\bh_u^{\rm H} \bw \right|^2 + \sigma^2\right) - 2\Re\{u_u^* \bh_u^{\rm H}\bw \} + 1, $$
\begin{equation*}
	\begin{split}
		F_e(\bw,\bw_{\rm jam},u_e)=&|u_e|^2 \left(\left|\bh_e^{\rm H} \bw \right|^2 +\left|\bh_e^{\rm H} \bw_{\rm jam} \right|^2 + \sigma_e^2\right) \\
		&- 2\Re\{j u_e^* \bh_e^{\rm H}\bw \} + 1.
	\end{split}
\end{equation*}

The solutions to $u_u,\rho_u,u_e,\rho_e$ can be solved in closed forms:
\[
u_u^\star = \frac{\bh_u^{\rm H} \bw }{ \left| \bh_u^{\rm H}\bw \right|^2 + \sigma^2},~
\rho_u^\star = \left(1 - u_u^{\star,*} \bh_u^{\rm H}\bw \right)^{-1},
\]
\[
u_e^\star = \frac{j\bh_e^{\rm H} \bw }{ \left| \bh_e^{\rm H}\bw_{\rm jam} \right|^2 + \sigma_e^2},~
\rho_e^\star = \left(1 - u_e^{\star,*}j \bh_e^{\rm H}\bw \right)^{-1}.
\]

To address ${\cal CN}_{\rm th}$ and ${\cal CN}_{\rm SIC}$, we use the SCA-based algorithm to solve these constraints:
\begin{equation}
	\begin{split}
		&{\cal C}_{\rm th} \rightarrow 2\Re\{ \hat\bw^{\rm H} \bh_u\bh_u^{\rm H} \bw \} - \Re\{ \hat\bw^{\rm H} \bh_u\bh_u^{\rm H} \hat\bw \} \geq r_{\rm th} \sigma^2, \\
		&{\cal C}_{\rm SIC} \rightarrow 2\Re\{ \hat\bw_{\rm jam}^{\rm H} \bh_u\bh_u^{\rm H} \bw \} - \Re\{ \hat\bw_{\rm jam}^{\rm H} \bh_u\bh_u^{\rm H} \hat\bw_{\rm jam} \} \geq \left|\bh_u^{\rm H}\bw \right|^2, 
	\end{split}
\end{equation}
where $\hat\bw_{\rm jam}$ and $\hat\bw$ denote the optimal solution in previous iteration loop.

After variation, \textbf{P2} is convex and can be solved by the existing toolbox such as CVX.

\subsection{Phase Shift Matrix Optimization for XL-RIS}
Given the precoding vectors $\bw$ and $\bw_{\rm jam}$ at the BS, \textbf{P1} can be expressed as
\begin{equation}
	\begin{split}
	\textbf{P3}:\max_{\btheta}~g(\btheta)~~{\rm s.t.}~{\cal C}_{\rm th},{\cal C}_{\rm SIC},{\cal C}_{\rm RIS}.
	\end{split}
\end{equation}

To solve \textbf{P3}, we rewrite it in the augmented Lagrangian form 
\begin{equation}
	\begin{split}
		\min_{\btheta,\widetilde\btheta}~g(\btheta)+\mu\left\| \widetilde\btheta - \btheta + \bm\nu \right\|_2^2 ~~{\rm s.t.}~{\cal C}_{\rm th},{\cal C}_{\rm SIC},{\cal C}_{\rm RIS},
	\end{split}
\end{equation}
where $\mu > 0$ and $\bm\nu\in \mathbb{C}^{N^2\times 1}$.

This non-convex problem is solved by ADMM with three iterative steps:
\begin{equation}
	\begin{split}
		&\btheta^{m+1} = \arg\min_{\btheta}~g(\btheta)+\mu\left\| \widetilde\btheta^m - \btheta + \bm\nu^m \right\|_2^2 ~{\rm s.t.}~{\cal C}_{\rm th},{\cal C}_{\rm SIC}, \\
		&\widetilde\btheta^{m+1} = \arg\min_{\widetilde\btheta}\left\| \widetilde\btheta - \btheta^{m+1} + \bm\nu^m \right\|_2^2 ~{\rm s.t.}~{\cal C}_{\rm RIS}, \\
		&\bm\nu^{m+1} = \bm\nu^m - \btheta^{m+1} + \widetilde\btheta^{m+1}.
	\end{split}
\end{equation}

As the same in Section III-B, we apply the WMMSE~\cite{Zhang2024ARIS} and SCA-based algorithms to convert step i) into a convex form, which can be solved via CVX.
To solve the subproblem ii), a direct projection method is proposed
\begin{equation} 
	\widetilde\btheta^{m+1} = 
	\begin{cases}
		\btheta^{m+1} - \bm\nu^m,~{\rm if}~ \left| \theta_n^{m+1} - \nu_n^m \right|=1, \\
		\frac{\btheta^{m+1} - \bm\nu^m}{\left| \btheta^{m+1} - \bm\nu^m \right|},~~{\rm otherwise}.
	\end{cases}
\end{equation}

However, implementing continuous phase shifts is impractical. 
Hence, we adopt a line search strategy for optimizing discrete phase shifts. For b-bit quantization. 
There are $2^b$ discrete phase levels ${\cal F} \in \{ -\pi,2\pi/2^b-\pi,\dots,(2^b-1)2\pi/2^b-\pi \} $~\cite{Zhang2024Prac_RIS}. 
The line search-based discrete phase optimization can be expressed as
\begin{equation}
	\begin{split}
		\widetilde\theta_n^{m+1} = \arg\min_{\widetilde\theta_n}~\left\| \widetilde\theta_n - \theta_n^{m+1} + \nu_n^{m} \right\|_2^2~~{\rm s.t.}~\widetilde\theta_n \in {\cal F}.
	\end{split}
\end{equation}

\subsection{Computational Complexity Analysis}
The main computational complexity of the algorithm lies in solving \textbf{P2} and \textbf{P3}. 
The complexity of solving the SOCP of \textbf{P2} is ${\cal O}(M^3K^3)$, and the complexity of solving the ADMM corresponding to \textbf{P3} is ${\cal O}(N^3)$.

\begin{figure}[t]	
	\centering \includegraphics[width=\linewidth]{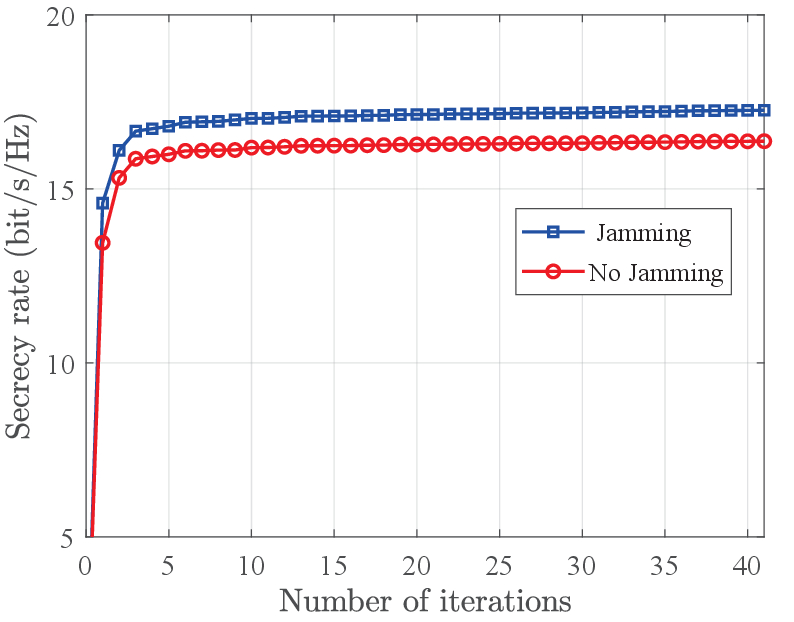}
	\caption{Convergence behaviors of the proposed algorithm.}
	\label{fig:iter}
\end{figure}
\begin{figure}[t!]
	\centering
	\subfloat[The distance from 5m to 13m ]{\includegraphics[width=\columnwidth]{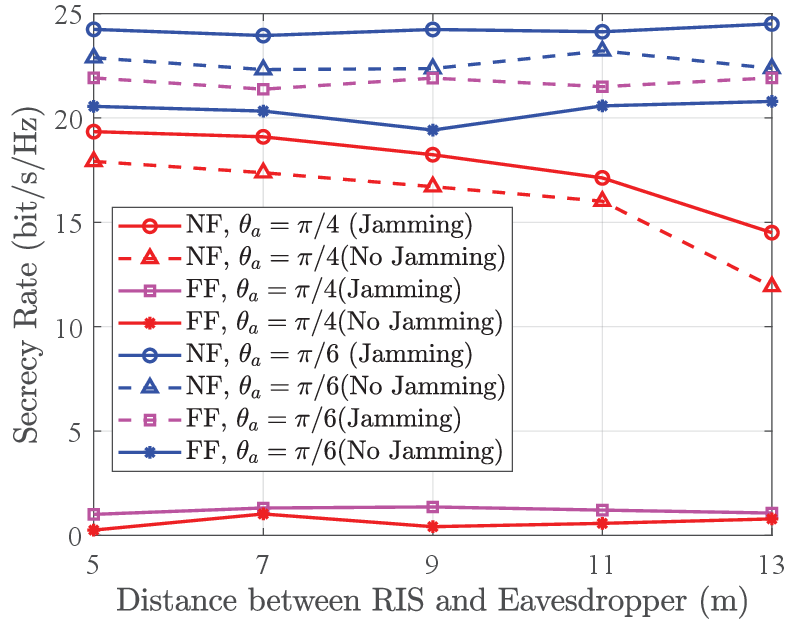}%
	}
	\hfil
	\subfloat[The distance from 17m to 25m ]{\includegraphics[width=\columnwidth]{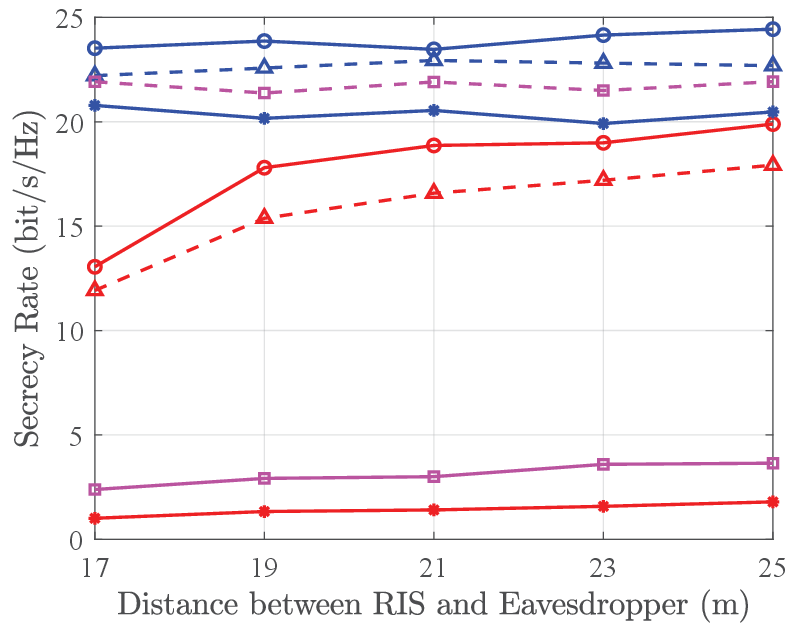}%
	}
	\caption{Secrecy rate versus distance between the XL-RIS and the eavesdropper.}
	\label{fig:distance}
\end{figure}
\section{Simulation Results}
In this section, simulation results are presented to show the performance of the pro-posed XL-RIS-aided near-field secure communication with artificial jamming. 
The XL-RIS is placed on the yz-plane, with its midpoint located at (0, 0, 0). 
The BS is positioned at (100, -100, 0). 
Both the legitimate user and eavesdropper are located on the $x-y$-plane. 
For convenience, their positions use a polar coordinate system. 
The eavesdropper is located at $(10{\rm m},\pi/4)$ and the legitimate user is located at $(15{\rm m},\pi/4)$. 
The default parameters are set as $M=8$, $N=64\times 8$, $\sigma^2=\sigma_e^2=-80{\rm dBm}$, $r_{\rm th}=1$, $f=10{\rm GHz}$, $P_{\rm max}=10{\rm dBm}$.
The Rayleigh distance is $d_R=62.4$m. 
The far-field Rician channel fading $\bG = \sqrt{\beta_0 d_{\rm BR}^{-\alpha_{\rm BR}}} \left(\sqrt{\frac{\kappa_{\rm BR}}{\kappa_{\rm BR}+1}} \bG_L + \sqrt{\frac{1}{\kappa_{\rm BR}+1}} \bG_S   \right) $, where $\beta_0 = -30$dB, $\alpha_{\rm BR}=2.2$ and $\kappa_{\rm BR}=3$dB.
$\bG_L \in \mathbb{C}^{N\times M}$ is the line-of-sight (LoS) component.
$\bG_S \sim {\cal CN}(0,\bI)$ is the scattering component. For the near-field channel, a single-path LoS channel is used. 
The channel gain is $\beta_1=(1+j)\zeta$, where $\zeta\sim {\cal N}(1,0.1^2)$.

\begin{figure}[t]	
	\centering \includegraphics[width=\linewidth]{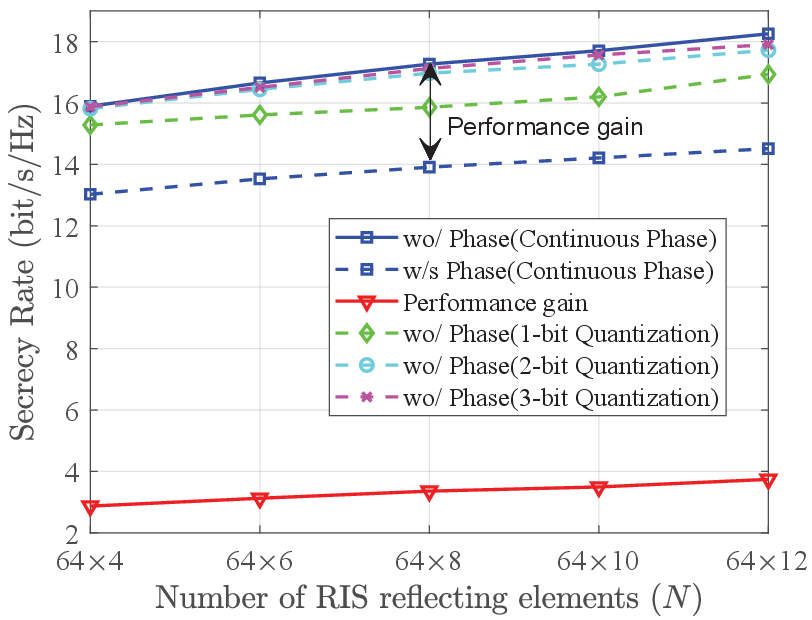}
	\caption{Secrecy rate versus number of XL-RIS’s elements $N$.}
	\label{fig:ris_gain}
\end{figure}

Fig.~\ref{fig:iter} shows the convergence performance of the proposed algorithm. It can be seen that the algorithm achieves convergence within 40 iterations. Furthermore, the scheme with artificial jamming has a higher secrecy rate than the scheme without it.

Fig.~\ref{fig:distance} plots the secrecy performance against the distance between the XL-RIS and the eavesdropper for different angles. When the legitimate user and the eaves-dropper share the same azimuth angle of $\pi/4$, the proposed near-field (NF) scheme can always
achieve a satisfactory secrecy rate. Furthermore, the secrecy rate decreases to zero as the location of the eavesdropper moves from $(5{\rm m},\pi/4)$ to the legitimate user loca-tion $(15{\rm m},\pi/4)$, and then increases as the location of the eavesdropper further moves far away from the legitimate user. This indicates that the closer to the XL-RIS loca-tion, the stronger beam diffraction ability of the XL-RIS. In contrast, the secrecy rate is almost zero in the far-field (FF) communication system. This comparison demon-strates that at the same angular direction, the near-field model can effectively ensure secure communication, while the far-field model cannot. In addition, when the eavesdropper’s azimuth angle is different from that of the legitimate user (e.g., set to $\pi/6$), the variation in the secrecy rate with respect to the distance becomes diminished.

For comparison, we consider the scheme with stochastic (`w/s') phase shift $\btheta$ at the XL-RIS in Fig.~\ref{fig:ris_gain}. 
The secrecy rate with the optimized (‘wo/’) phase-shift scheme is significantly higher than that of the stochastic case. The performance gap becomes more bigger as the number of $N$ increases. Furthermore, with 2- and 3-bit quantization, the secrecy performance is very close to the continuous case, demonstrating that discrete phase control can meet practical deployment needs while reducing hardware cost.

\section{Conclusion}
In this paper, we have studied the XL-RIS-aided near-field secure communication via artificial jamming. The secrecy rate is maximized by jointly optimizing the beamforming vector at BS and the phase shift matrix at XL-RIS. A low-complexity AO algorithm is proposed to solve the problem. Furthermore, the proposed algorithm can accommodate discrete phase optimization for the XL-RIS, which reduces hardware costs. Simulation results show that when the eavesdropper tends to be located in security-blind zones to wiretap the legitimate user, the scheme can still ensure secure transmission. For future works, we plan to consider the impact of imperfect CSI.

\bibliographystyle{IEEEtran}
\bibliography{refs}

\end{document}

%% file: sym_marco.tex
\newcommand\bw{\ensuremath{{\bm w}}}

\newcommand\bx{\ensuremath{{\bm x}}}

\newcommand\bG{\ensuremath{{\bold G}}}

\newcommand\bh{\ensuremath{{\bm h}}}

\newcommand\bz{\ensuremath{{\bm z}}}

\newcommand\ba{\ensuremath{{\bm a}}}

\newcommand\bb{\ensuremath{{\bm b}}}

\newcommand\btheta{\ensuremath{{\bm \theta}}}
\newcommand\bTheta{\ensuremath{{\bm \Theta}}}

\newcommand{\diag}{\mathrm{diag}}

\newcommand\br{\ensuremath{{\bm r}}}

\newcommand{\bI}{{\bold I}}